\documentclass[twocolumn,twoside,slac_two]{revtex4}
\usepackage{graphicx}
\usepackage{fancyhdr}
\usepackage{hyperref}
\usepackage{siunitx}
\usepackage{subfigure}
\hypersetup{
    colorlinks=true,
    urlcolor=magenta,
}
\begin{document}

\title{EUSO-TA fluorescence detector}

\author{F. Bisconti (for the JEM-EUSO Collaboration)}
\affiliation{Karlsruher Institut f\"ur Technologie (KIT), Institut f\"ur Kernphysik (IKP), Hermann-von-Helmholtz-Platz 1, 76344 Eggenstein-Leopoldshafen}
\begin{abstract}
EUSO-TA is a pathfinder experiment for the space based JEM-EUSO mission for the detection of ultra-high energy cosmic rays. EUSO-TA is an high-resolution fluorescence telescope installed in front of the Black Rock Mesa fluorescence detectors of the Telescope Array (TA) experiment, in Utah (USA). At the TA site, a Central Laser Facility is installed for calibration purposes, since it emits laser beams with known energy and geometry. EUSO-TA consists of two 1 $\mbox{m}^2$ Fresnel lenses, with a field of view of \ang{10.5} that focus the light on a Photo Detector Module (PDM). The PDM currently consists of 36 Hamamatsu Multi-Anode Photo-Multipliers Tubes (MAPMTs) with 64 channels each. Front-end readout is performed by 36~ASICS, with two FPGA boards that send the data to a CPU and a storage system. The detector was installed in February 2015. Tests using the mentioned light sources have been performed and observations of cosmic ray events, as well as those of stars with different magnitude and color index have been done. The data acquisition is triggered by TA fluorescence detectors, although a self-trigger algorithm is currently in the last phases of development and test. With TA, thanks to its large field of view and the surface detectors, the cosmic ray shower events are reconstructed and the parameters are used to perform simulations of the response of EUSO-TA detector using \textit{Offline}. Simulations of the detected events are compared with data and the results are shown in this work.
\end{abstract}

\maketitle

\thispagestyle{fancy}

\section{Introduction}
\begin{figure}[!htbp]
	\includegraphics[width=8cm]{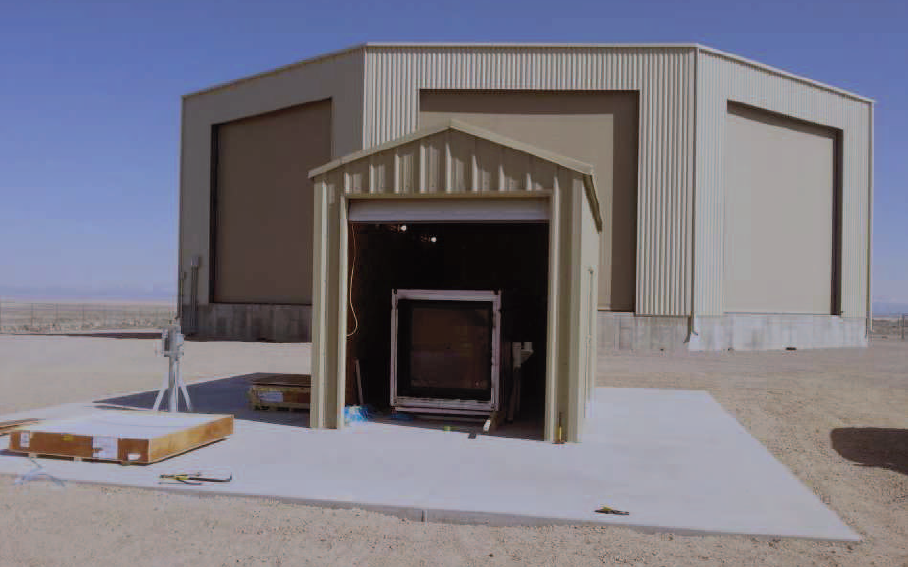}
	\caption{\footnotesize The EUSO-TA dome in front of the BRM-FDs station.}
	\label{fig:euso-ta-site}
\end{figure}

EUSO-TA \cite{adams_2015} is a pathfinder detector of the JEM-EUSO project \cite{adams_2014}. The aim of JEM-EUSO is to detect the rare ultra-high energy cosmic rays (UHECRs) from space, observing the UV fluorescence light emitted by cosmic ray extensive air showers (EASs) through the atmosphere, which is used as an huge calorimeter.\\
The purpose of the EUSO-TA experiment is to validate some aspects of the observation principle of JEM-EUSO detecting EASs from ground. Indeed, it is located at the Telescope Array (TA) \cite{fukushima_2003} site in Utah (USA), in front of the Black Rock Mesa fluorescence detectors (BRM-FDs) station \cite{tokuno_2012}, both visible in Fig.~\ref{fig:euso-ta-site}.
\section{Design}
\begin{figure}[!htbp]
	\centering
	\subfigure{\includegraphics[height=3.7cm]{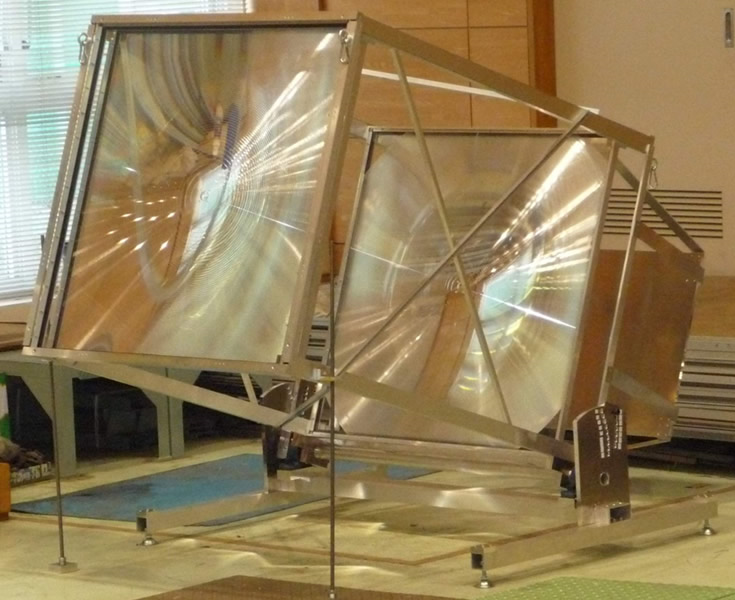}}
	\subfigure
	{\includegraphics[height=3.7cm]{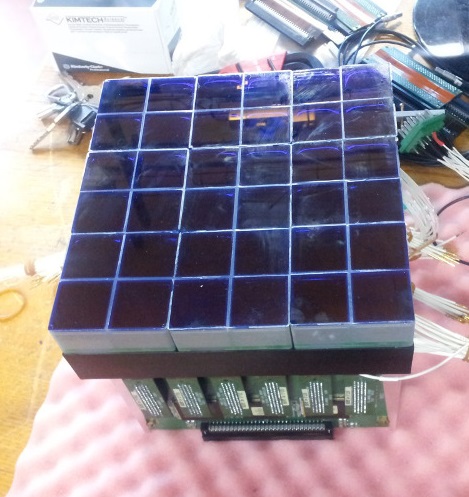}}
	\caption{\footnotesize EUSO-TA optical system (left) and PDM (right).}
	\label{fig:euso-ta-design}
\end{figure}

EUSO-TA has the same basic design used also in EUSO-Balloon \cite{ballmoos_icrc2015} and EUSO-SPB \cite{wiencke_icrc2015} telescopes, other two JEM-EUSO pathfinders dedicated to the validation of the detector design. \\
The optical system is made by two $1\times1\,\mbox{m}^2$ flat lenses, 8~mm thick each. The design of the optical system is visible in Fig.~\ref{fig:euso-ta-design} left. The front side of the front lens and the back side of the rear lens have a plane surface, which is useful for EUSO-TA in order to clean the dust easily, since it is placed in a desert area. The back side of the front lens and the front side of the rear lens have a Fresnel structure. They are made of UV grade polymethyl-methacrylate (PMMA) and details about their manufacture can be found in \cite{hachisu_icrc2013}. The optical system allows to EUSO-TA a Field Of View (FOV) of \ang{10.5}.\\
The EUSO-TA focal surface has a concave shape. It consists of one Photo Detector Module (PDM), which is a $17\times17$~cm active surface composed by an array of $6\times6$ Hamamatsu R11265-M64 MAPMTs. Groups of $2\times2$ MAPMTs form the Elementary Cells (ECs). The PDM is shown in Fig.~\ref{fig:euso-ta-design} right. \\
Each MAPMT is a matrix of $8\times8$~pixels with 2.88 mm side for a total of 2304 pixels over the whole PDM. An UV transmitting band pass filter (Schott BG3) in the range 290~nm-430~nm is glued on top.
Each pixel has a gain (electron multiplication ratio) of more than $10^6$ which allows single photon counting. The total photon detection efficiency has a maximum of 35\%.\\
\section{Operation}
As all the fluorescence cosmic ray detectors, EUSO-TA works in night time and  possibly in clear sky conditions, to reduce effects of atmospheric and cloud attenuation. The elevation of the instrument can be manually changed from \ang{0} to \ang{25}, whereas the azimuth is fixed to \ang{53} from North counterclockwise, being the telescope pointing to the CLF direction.\\
One advantage of the EUSO-TA location is the possibility to use one of the systems developed by the TA collaboration for calibration purposes: the Central Laser Facility (CLF) \cite{udo_2007}. The CLF is a laser facility 21~km distant from BRM-FDs which shoots every half an hour 300 laser pulses vertically with a frequency of 10~Hz at a known energy, with a typical energy of \~4~mJ that can be adjusted. The light scattered from the beam is equivalent to that produced by an equally distant UHECR with energy of $~10^{19.5}$~eV. \\
The second advantage is to have the possibility to acquire data in coincidence with BRM-FDs (\textit{external trigger}). Indeed, in case of a trigger signal received by BRM-FDs, EUSO-TA starts to save the data from the buffer in packets of 128 Gate Time Units (GTUs), which otherwise would be overwritten (one GTU corresponds to $2.5\,\mu\mbox{s}$, and represents the time resolution of the data acquisition). This is an important point since this acquisition mode allows to know that EUSO-TA could have detected an event and, from the shower reconstruction made by the TA experiment, the nature of the event (i.e. direction, distance, energy etc.) would be then known and used for further analysis. However, since the FOV of TA is about 30 times the FOV of EUSO-TA, the amount of data saved is larger than the actual data containing events, that makes the research of events more challenging.
Another acquisition mode consists in using an \textit{internal trigger}, also called \textit{first level trigger} (L1) \cite{bertaina_2016}, implemented for the JEM-EUSO experiment. The L1 principle consists in counting an excess of signals over background in groups of $3\times3$ pixels lasting more than a preset time. 
The internal trigger has been tested, but the current FPGA allows the implementation of the trigger logic only for a few ECs. For this reason, most of the data have been acquired with the external trigger.\\
\section{Analysis and simulations}
The detected events have been analyzed and simulated, but as an example, only the first detected event is discussed in detail in this contribution. It has been detected on the 13th May 2015 in external trigger mode and, since it was with about 2.5~km relatively close to the detector, it lasted 1 GTU.
\subsection{Overall event analysis}\label{sec:background}
\begin{figure}[h]
	\centering
	\includegraphics[width=8.2cm]{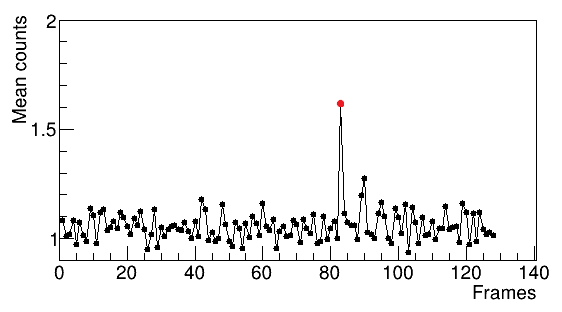}
	\caption{\footnotesize Sequence of 128~mean values of frames in the packet containing the detected event (red mark at frame number~83).}
	\label{fig:mean128_13may}
\end{figure}

A study on the variability of the background has been done to evaluate if the presence of CR event signal influences in an evident manner the mean of the pixels' content of a whole frame. For this purpose, 128~frames corresponding to the packet of GTUs containing the detected event have been taken into account. For each frame the mean value of the pixel contents has been calculated and the 128~mean values are plotted in Fig.~\ref{fig:mean128_13may}. It shows the sequence of 128~mean values and, with red mark, the frame with signal from the CR event is highlighted (frame number~83 in the graph). It is evident that the contribute of the CR event is easily visible over the base line of the background. Unfortunately, from the same study on other detected events it has been found that for events spread on more than one GTU, it is impossible to find the frames with the CR event since the contribution of just a segment of the whole event track has the same magnitude as the background fluctuations.
\subsection{Event simulation}
\begin{figure}[!htbp]
	\centering
	\subfigure[\protect Data]{
		\includegraphics[width=6.5cm,height=6.5cm]{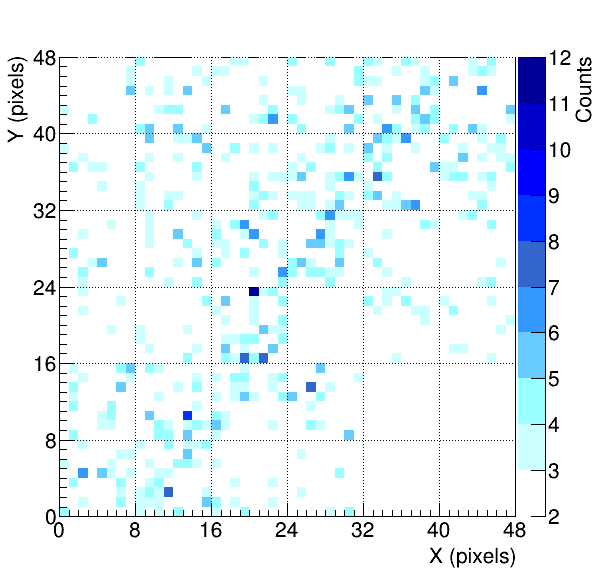}}
	\subfigure[\protect Simulation]{
		\includegraphics[width=6.5cm,height=6.5cm]{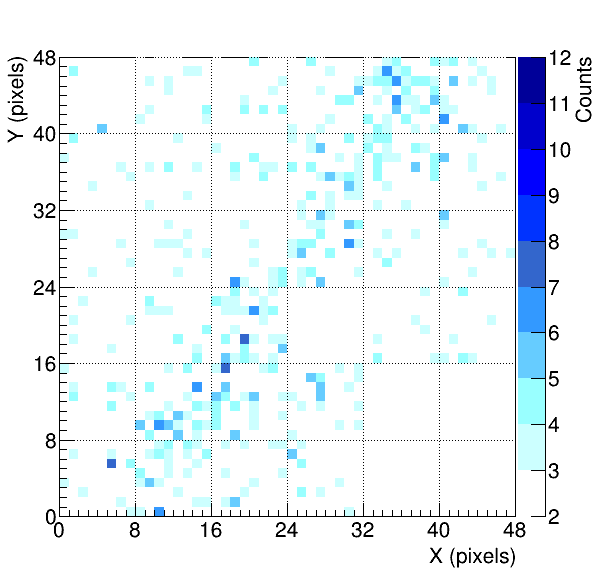}}
	\subfigure[\protect BRM-FDs event with EUSO-TA's FOV overlapped. Marker sizes are proportional to square root of charge and colors are corresponding to signal's peak timing (violet means earlier, red means later). Image by courtesy of the TA collaboration.]{
		\includegraphics[width=8.2cm]{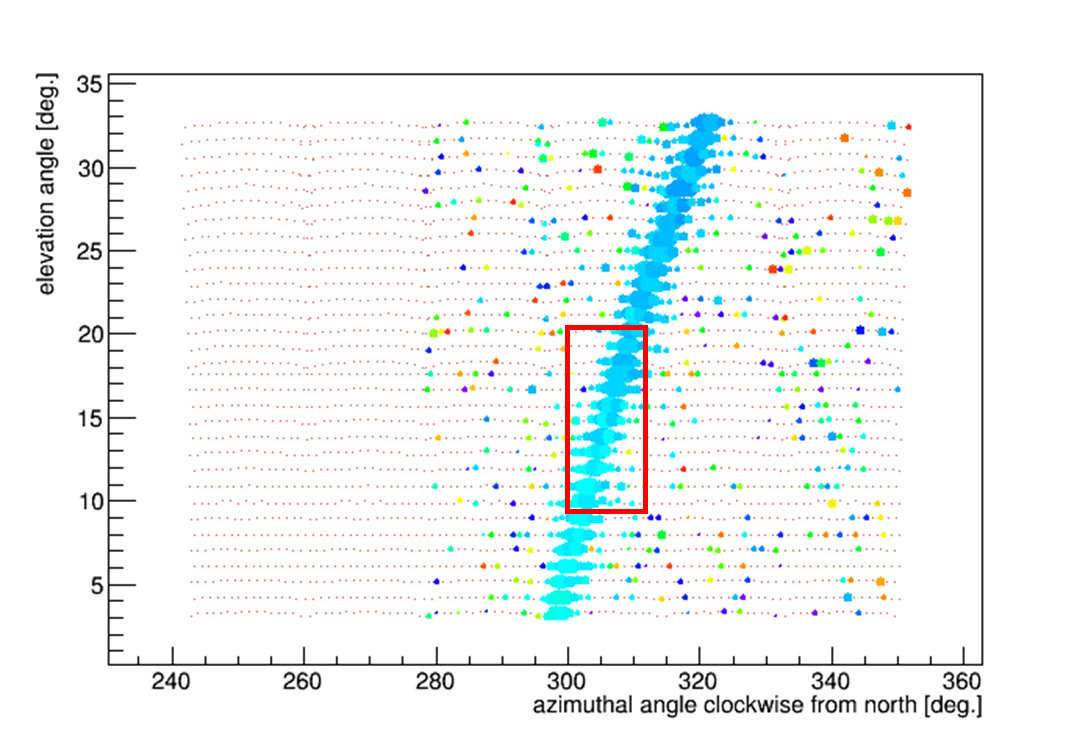}}
	\caption{
		\footnotesize Comparison between data (a) and simulation (b) of the event detected by EUSO-TA. The same event as detected by BRM-FDs with EUSO-TA's FOV overlapped (c).}
	\label{fig:may13_dat_sum}
\end{figure}

\begin{figure}[!htbp]
	\centering
	\includegraphics[width=6.5cm, height=5.6cm]{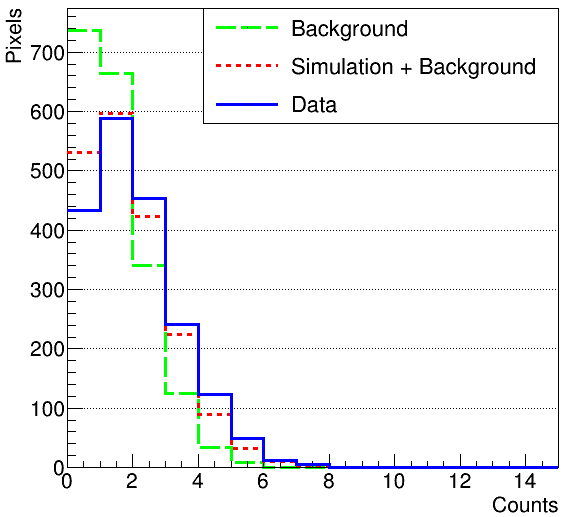}
	\caption{
		\footnotesize Comparison between the counts distributions of data,  sum of simulation and background and background.}
	\label{fig:may13_diff}
\end{figure}

Two simulation frameworks are used in the JEM-EUSO collaboration: EUSO-Offline \cite{paul_icrc2015} and ESAF \cite{fenu_icrc2011}. In this work, results of simulations made with EUSO-Offline are discussed. 
Improvements regarding the telescope and electronics simulators have been recently made in EUSO-Offline, that made the simulations properly comparable with the data. \\
Continuing the discussion on the 13th May 2015 event, we consider the comparison between the data and simulation.
Tab.~\ref{tab:13may_sim} shows the event reconstruction of this event performed by TA experiment, together with the elevation angle that EUSO-TA had during the data acquisition. Most of these information are required as input in the simulation and are marked with a star in the table: zenith and azimuth (counterclockwise from East) angles of the shower, energy of the primary particle, coordinates of the shower core on ground (in the UTM coordinate system, zone "12" band "S") and the elevation of EUSO-TA.
\begin{table}[h]
	\begin{center}
		\begin{tabular}{|l|l|}
			\hline
			\multicolumn{2}{|c|}{\textbf{Shower parameters}} \\
			\hline
			* Zenith                 & \ang{35}  \\
			* Azimuth                & \ang{7} \\
			* Energy                 & $10^{18}$ eV \\
			$R_p$ (impact param.)    & 2.5 km \\
			* Core coordinate        & 349976 m East, 4340200 m North\\
			Distance core            & 2.43 km \\
			Number of GTUs           & 1 \\
			\hline
			\hline
			\multicolumn{2}{|c|}{\textbf{EUSO-TA configuration}} \\
			\hline
			* Elevation              & \ang{15} \\
			\hline
		\end{tabular}
		\caption{Event parameters.}\label{tab:13may_sim}
	\end{center}
\end{table}\\
In Fig.~\ref{fig:may13_dat_sum} the comparison between the detected (a) and the simulated (b) event is shown, where both the frames show the number of counts on the full PDM. The simulation itself has been summed to a frame of background extrapolated from the data, from the same packet of GTUs where the event is saved. The white blocks correspond to not working MAPMTs, where correspondingly also the simulation has been masked. They look in good agreement from a visual point of view: the number of counts, the inclination and width of the tracks look similar. In (c), the same event as detected by BRM-FDs within BRM-FD's $51.2\,\mu \mbox{s}$ window is shown, with EUSO-TA's FOV overlapped. It is visible that the spatial resolutions are higher for EUSO-TA, although the FOV is quite smaller.
A quantitative comparison is done considering the counts distribution of the data, the simulation with background and the background itself, as shown in Fig.~\ref{fig:may13_diff}. The not working PMTs are not considered in the analysis. The distributions of data and simulation differ slightly. The difference probably resides, at least in part, in the impossibility to reproduce the exact background frame existing at the moment when the frame with the event was taken. But it is interesting to see how both the distributions start to show an excess respect to the background at $\mbox{counts} \geq 2$.
\section{Statistics}
In the year 2015 four data tacking campaigns have been done in coincidence with BRM-FDs, and one in 2016, for a total of about 140~hours. Up to now four events are considered surely detected. Two of them are related to close showers, lasting 1~GTU, while the other two, at further distances, last 2 and 3 GTUs. To give a reference, events with distance $R_p\geq2.5$~km should have an energy $E\geq10^{18}$~eV to be detected.\\
A study on all the events detected by BRM-FDs and passing inside the EUSO-TA FOV is ongoing. First, each event is simulated and, if visible in the simulation, it is searched in the data, with the help the timestamp of the corresponding TA event. A direct research of the events in EUSO-TA data just via the timestamp of the TA events is not efficient, because of a not constant time shift between the timestamps of the events of the two experiments and because of the different time resolutions. Currently, four more events found in the data are considered as candidate events, and further analysis is proceeding in order to ensure their classification.
The collected statistics is compatible with the expectations from preliminary analysis. 
\section{Conclusions}
EUSO-TA is currently active and is successfully collecting data. The Offline simulation framework has been improved and it is able to simulate quite well the events. Further analysis on the not detected events are ongoing to search for not yet seen events in EUSO-TA data.

\begin{acknowledgments}
\footnotesize 
The support received by TA collaboration is deeply acknowledged.\\
This work was partially supported by Basic Science Interdisciplinary 
Research Projects of RIKEN and JSPS KAKENHI Grant (22340063, 23340081, and 
24244042), by the Italian Ministry of Foreign Affairs, General Direction 
for the Cultural Promotion and Cooperation, by the 'Helmholtz Alliance 
for Astroparticle Physics HAP' funded by the Initiative and Networking Fund 
of the Helmholtz Association, Germany, by NASA award 11-APRA-0058 in the USA, and by Slovak Academy of Sciences 
MVTS JEM-EUSO as well as VEGA grant agency project 2/0076/13.
Russia is supported by the Russian Foundation for Basic Research 
Grant No 13-02-12175-ofi-m.
The Spanish Consortium involved in the JEM-EUSO Space Mission is funded by 
MICINN \& MINECO under the Space Program projects: AYA2009-06037-E/AYA, 
AYA-ESP2010-19082, AYA-ESP2011-29489-C03, AYA-ESP2012-39115-C03, 
AYA-ESP2013-47816-C4, MINECO/FEDER-UNAH13-4E-2741, CSD2009-00064 
(Consolider MULTIDARK) and by Comunidad de Madrid (CAM) under projects 
S2009/ESP-1496 \& S2013/ICE-2822.
\end{acknowledgments}

\bigskip 

\end{document}